\documentclass{JHEP3}

\newcommand{\be}{\begin{equation}}
\newcommand{\ee}{\end{equation}}
\def\bea{\begin{eqnarray}}
\def\eea{\end{eqnarray}}

 \def\be{\begin{equation}}
\def\ee{\end{equation}}
\def\bea{\begin{eqnarray}}
\def\eea{\end{eqnarray}}
\def\lesssim{\mathrel{\hbox{\rlap{\hbox{\lower4pt\hbox{$\sim$}}}\hbox{$<$}}}}
\def\gtrsim{\mathrel{\hbox{\rlap{\hbox{\lower4pt\hbox{$\sim$}}}\hbox{$>$}}}}

\title{Issues in Type IIA Uplifting}
 \author{ Renata Kallosh$^a$\footnote{\mbox{
Email: {\tt kallosh@stanford.edu} }} and Masoud
Soroush$^{a,b}$\footnote{\mbox{ Email: {\tt soroush@stanford.edu} }}
\\
$^a$Department of Physics, Stanford University, Stanford, CA 94305,
USA \\
$^b$SLAC, Stanford University, Stanford, CA 94309, USA}
\received{\today} 
 \preprint{SU-ITP-2006-34 \\
  SLAC-PUB-12251\\ December 7, 2006}
\abstract{Moduli stabilization in the type IIA massive string theory
so far was achieved only in the AdS vacua. The uplifting to dS vacua
has not been performed as yet: neither the  analogs of type IIB
anti-D3 brane at the tip of the conifold, nor the appropriate
D-terms have been identified. The hope was recently expressed that
the F-term uplifting may work. We investigate this possibility in
the context of a simplified version of the type IIA model developed
in hep-th/0505160 and find that the F-term does not uplift the AdS
vacua to  dS vacua with  positive CC. Thus it remains a challenging
task to find phenomenologically acceptable vacua in the type IIA
string theory.}
\begin{document}
\section{Introduction}

Models with all moduli stabilized in the massive type IIA string
theory are relatively simple: they engage all possible fluxes, no
non-perturbative  effects from the perspective of the massive
ten-dimensional type IIA supergravity are required
\cite{Derendinger:2004jn,DeWolfe:2005uu}. A model of particular
interest which we will study here, is the type IIA compactification
on the $T^6/\mathbb{Z}_{3}^{2}$ orientifold studied in
\cite{DeWolfe:2005uu}. It solves the equations of motion of the 10D
type IIA massive supergravity with account of local O6 and D6
sources and all possible fluxes. An interesting feature of this
model is that the stabilized moduli depend on a particular integer
$N$ related to certain fluxes. $N$ can be arbitrarily large, in
contrast to other known models with stabilized moduli where fluxes
are bounded, see for example a review on this in
\cite{Douglas:2006es}. The action for this model, as well as for
more general models in this class, was constructed in
\cite{Grimm:2004ua}. More recently, various stringy aspects of this and related
models were studied in \cite{Acharya:2006ne}.

The uplifting to dS vacua for these models have not been performed
so far; neither the  analogs of the type IIB anti-D3 brane at the
tip of the conifold nor the appropriate D-terms have been
identified. There are models in type IIA string theory in which dS vacua have been found \cite{Saueressig:2005es}. In these models there is either a full hypermultiplet including dilaton-axion is stabilized or in the orientifold case, only the dilaton-axion is stabilized. The stabilization is achieved via background fluxes and membrane instanton corrections, K\"{a}hler moduli are assumed to be integrated out. So far this approach has not been applied yet to the model \cite{DeWolfe:2005uu} which we study here.  Finally, the  recently improved version of the K\"{a}hler
uplifting \cite{Westphal:2005yz} was not yet applied to the model in
\cite{DeWolfe:2005uu}, so we have to wait before we know if it
works.

On the other hand, recently there was a significant development with
the F-term uplifting  in combining the  KKLT model
\cite{Kachru:2003aw}  with models which separately have
non-supersymmetric vacua with positive energy
\cite{Gomez-Reino:2006dk, Dudas:2006gr,Kallosh:2006dv}. The fields
in these models are expected to originate from the open string
sector in intersecting brane models, however, the details still have
to be worked out. If the uplifting model has small vev's comparative
to Planckian scales, the leading part of the F-term uplifting does
not depend on the details of the added model. Perhaps the simplest
of these models is the O'KKLT model \cite{Kallosh:2006dv} combining
the O'Raifeartaigh model \cite{O'Raifeartaigh:1975pr} with the
Coleman-Weinberg corrections and the KKLT model.

The generic feature of F-term uplifting models is that
\be K=
K_A+K_B \ , \qquad W= W_A +W_B \ . \ee Here the model $A$ is
gravitational and the model $B$ is close to global susy: all
dimensionful quantities in model $B$ are much smaller than the
Planckian scale. In such case \be V_{total}\approx  V_{A}+ e^{K_A}
V_B +... \ee The examples were proposed in
\cite{Dudas:2006gr,Kallosh:2006dv}. In string theory setting the
fields in $A$ originate from the closed string sector and the fields
in  $B$ originate from the open string sector.

It was suggested in \cite{Dudas:2006gr} that the new models of
F-term uplifting can be applied in type IIA models of moduli
stabilization \cite{Derendinger:2004jn,DeWolfe:2005uu}. Here we will
study this problem for the model of the massive type IIA string
theory  with all moduli stabilized in \cite{DeWolfe:2005uu}. There
are 13 complex scalars in this model, therefore the potential is
extremely complicated. 3 untwisted moduli $t_i, i=1,2,3$ represent
the volume-axion for each of the  2-tori $T^6= T^2\times T^2\times
T^2$. There are 9 blow-up moduli, $t_{B_A}, A=1, ..., 9$, associated
with 9 singular orbifold points. Finally, there is a complex
dilaton-axion field $N_0$, coming from the quaternionic part of the
geometry, only half of quaternion components, forming a complex
field,  remains after the orientifolding.

We will consider a somewhat simplified version of this model, in
which the choice of fluxes is such that i) the stabilized axions
are equal to zero, $\rm Re \, t_i =0 , \rm Re \, t_{B_A}=0$ and ii)
the untwisted volumes    are significantly larger than blow-ups $\rm
Im \, t_i\gg \rm Im \, t_{B_A}$. In the regime of the validity of
the supergravity approximation, this simplified model captures the
features of the most general class of models. Therefore it is
plausible that the situation with uplifting in this simplified model
will also tell us what is going to happen in  the general case.
However, the most general case may need a separate careful
investigation, which may  prove that the outcome of the F-term
uplifting is the same or better than in the simplified case.

\section{Type IIA Compactification on $T^6/Z_{3}^{2}$ Orientifold}

{\underline {The field content}:}

\

Since $h^{2,1}(T^{6}/Z_{3}^{2})=0$, it has no complex structure
modulus field. But it has the following K\"{a}hler moduli fields: it
has 3 K\"{a}hler moduli $\{t_{i}, i=1,2,3\}$, each associated with
one torus $T^{2}$. Also, it has 9 K\"{a}hler moduli $\{t_{B_A},
A=1,2,\cdots 9\}$ associated with 9 singularities of the orbifold
(blow-up modes). Further,  the  dilaton $e^{-\phi}$ and its axionic
partner $\xi$ are coming from the complex field $N_0$.

\textit{In short, this model has 12 complex fields (K\"{a}hler
moduli) $\{t_{i},t_{B_A}; i=1,2,3,\ A=1,2,\cdots 9\}$, and an
additional complex scalar, $N_0$, coming from the universal
hypermultiplet after the orientifolding}.

\

\noindent {\underline {The fluxes}:}

\

Since there are no complex structure moduli, we do not have any flux
which belongs to $H^{2,1}$. The only surviving part of the $H$ flux
comes from $H^{3,0}$. Therefore, as far as the $H$ flux is
concerned, we have \textit{only one} flux which we denote it by $p$
(which is a constant). We also have RR-fluxes which come from
$F_{0}$, $F_{2}$, $F_{4}$, and $F_{6}$ RR-forms. Let us parameterize
the fluxes (after integration of $p$-forms on $p$-cycles) by:
\begin{eqnarray}\label{a1}
&F_{0} \Longleftrightarrow  m_{0} \\
&F_{2} \Longleftrightarrow  \{m_{i},n_{A};\ i=1,2,3\ , A=1,2\cdots
9\}\\
&F_{4} \Longleftrightarrow  \{e_{i}, f_{A};\ i=1,2,3\ , A=1,2\cdots
9\}\\
&F_{6} \Longleftrightarrow  e_{0}
\end{eqnarray}
\textit{In short, $\{p,m_{0},e_{0},m_{i},e_{i},n_{A},f_{A};\
i=1,2,3\ ,A=1,2\cdots 9\}$ parameterize all fluxes in the most
general case for this orbifold}.

\

\noindent {\underline {The K\"{a}hler potential}:}

\

The components of intersection form of this orientifold are: $
\kappa_{ijk}=\kappa|\epsilon_{ijk}|\ ,\ \kappa_{AAA}=\beta\ , $ and
all other components vanish. The K\"{a}hler potential includes two
parts: The part which depend on the K\"{a}hler moduli:
\begin{eqnarray}\label{a2}
K^{K}=-\log\Big(-\frac{4}{3}(6\kappa
v_{1}v_{2}v_{3}+\beta\sum_{A=1}^{9}v_{bA}^{3})\Big)\ ,
\end{eqnarray}
in which we have rewritten each complex field in terms of its real
and imaginary parts: $t_{i}=b_{i}+iv_{i}$,
$t_{B_A}=b_{B_A}+iv_{B_A}$, and the part which depends on $H$ flux:
\begin{eqnarray}\label{a4}
K^{Q}=4D= {e^{4\phi}\over \rm {vol}^2}\ , \qquad \rm vol= \kappa
\upsilon_1  \upsilon_2    \upsilon_3
\end{eqnarray}
The total K\"{a}hler potential is $K=K^{K}+K^{Q}$.

\

\noindent  {\underline {The superpotential}:}

\

The total superpotential is given by  $W=W^{Q}+W^{K}$, where by
$W^{Q}=- 2 p N_{0}$ and
\begin{eqnarray}\label{a5}
W^{K}=e_{0}+e_{i}t_{i}+f_{A}t_{bA}+\frac{1}{2}\kappa|\epsilon_{ijk}|m_{i}t_{j}t_{k}+
\frac{1}{2}\beta\sum_{A=1}^{9}n_{A}t_{B_A}^{2}-m_{0}\kappa
t_{1}t_{2}t_{3}-\frac{m_{0}}{6}\beta\sum_{A=1}^{9}t_{bA}^{3}\ .
\end{eqnarray}

\


\noindent  {\underline {The supersymmetric solutions:}

\

The supersymmetric solutions are given by (we denote the solutions
by a star in order to keep things clear)
\begin{eqnarray}\label{a6}
t^{*}_{i}=\frac{m_{i}}{m_{0}}+\frac{i}{|\hat{e}_{i}|}\sqrt{\frac{-5\hat{e}_{1}\hat{e}_{2}\hat{e}_{3}}
{3m_{0}\kappa}}\ ,\ \ \ \mbox{where\ }\
\hat{e}_{i}=e_{i}+\frac{\kappa|\epsilon_{ijk}|m_{j}m_{k}}{2m_{0}}\ ,
\end{eqnarray}
\begin{eqnarray}\label{a7}
t^{*}_{B_A}=\frac{n_{A}}{m_{0}}-i\sqrt{\frac{-10\hat{f}_{A}}{3\beta
m_{0}}}\ , \ \mbox{where\ }\ \hat{f}_{A}=f_{A}+\frac{\beta
n_{A}^{2}}{2m_{0}}\ ,
\end{eqnarray}
and for the dilaton, we have
\begin{eqnarray}\label{a8}
e^{-\phi^{*}}=-\frac{4\sqrt{3}}{15}\frac{m_{0}}{p}\Bigg[\frac{10}{|m_{0}|}
\sqrt{\frac{-5\hat{e}_{1}\hat{e}_{2}\hat{e}_{3}}
{3m_{0}\kappa}}+\beta\sum_{A=1}^{9}\Big(\frac{-10\hat{f}_{A}}{3\beta
m_{0}}\Big)^{3/2}\Bigg]^{1/2}\ ,
\end{eqnarray}
and finally for the axion $\xi$, we have
\begin{eqnarray}\label{a9}
\xi^{*}=\frac{1}{p}\Big(e_{0}+\frac{e_{i}m_{i}+f_{A}n_{A}}{m_{0}}+\frac{6\kappa
m_{1}m_{2}m_{3}+\beta\sum_{A}n_{A}^{3}}{3m_{0}^{2}}\Big)\ .
\end{eqnarray}

We make two simplifications:

1. We choose  fluxes $F_2=0$, so that $m_i= n_A=0$. This leads to
$\hat e^i= e^i$ and $\hat f_A= f_A$ and
\begin{eqnarray}\label{a6}
t^{*}_{i}\Rightarrow\frac{i}{|{e}_{i}|}\sqrt{\frac{-5{e}_{1}{e}_{2}{e}_{3}}
{3m_{0}\kappa}} \ , \qquad t^{*}_{BA}\Rightarrow
-i\sqrt{\frac{-10{f}_{A}}{3\beta m_{0}}}\ .
\end{eqnarray}
For the dilaton  we have
\begin{eqnarray}\label{a8}
e^{-\phi^{*}}\Rightarrow
-\frac{4\sqrt{3}}{15}\frac{m_{0}}{p}\Bigg[\frac{10}{|m_{0}|}
\sqrt{\frac{-5{e}_{1}{e}_{2}{e}_{3}}
{3m_{0}\kappa}}+\beta\sum_{A=1}^{9}\Big(\frac{-10{f}_{A}}{3\beta
m_{0}}\Big)^{3/2}\Bigg]^{1/2}\ ,
\end{eqnarray}
and  for the axion $\xi$  we have
\begin{eqnarray}\label{a9}
\xi^{*}\Rightarrow\frac{1}{p}\Big(e_{0}\Big)\ .
\end{eqnarray}

2. We assume that fluxes on non-blow-up modes dominate the string coupling,\\
 $|e_i|\gg |f_A|\gg |m_0|$. This leads to an approximate equation for the stabilized dilaton,
\begin{eqnarray}\label{a8}
e^{-\phi^{*}}\Rightarrow
-\frac{4\sqrt{3}}{15}\frac{m_{0}}{p}\Bigg[\frac{10}{|m_{0}|}
\sqrt{\frac{-5{e}_{1}{e}_{2}{e}_{3}} {3m_{0}\kappa}}\Bigg]^{1/2}\ .
\end{eqnarray}


\section{F-term uplifting of the simplified model}

The total potential is given by \be V_{total}\approx V_{IIA} +
e^{K_{IIA}} V_{0} +... \ , \ee where $V_0$ is the constant value of
the potential from the $B$ sector. The uplifting term is given by
the following expression: \be V_{up}\approx e^{K_{IIA}} V_0\ ,
\qquad
  e^{K_{IIA}}= e^{K^K+K^Q} = {4D\over \rm
vol} \ . \ee
 If the $B$ sector is given by the  quantum
corrected  O'Raifeartaigh model  \cite{Kallosh:2006dv},
\cite{O'Raifeartaigh:1975pr}, we have   $W_{O'}=-\mu^2 X$ and
$K_{O'}= X\bar X- {(X\bar X)^2\over \Lambda^2}$, so that at the
minimum $e^{K_{O'}}\approx 1$ and $V_{O'}=V_0 \approx |{\partial W
\over \partial X}|^2= \mu^4$.

It is easy to study the uplifting in the simplified model since the
potential depends only on   2 variables. As shown in
\cite{DeWolfe:2005uu}, \be {1\over \lambda} V_{IIA} =  \Bigg[
{1\over 2} g^4 r^6 -\sqrt 2 g^3 + {1\over 4} {g^2\over r^6} +
{3\over 2} {g^4\over r^2}\Bigg]. \ee Here  $\upsilon_i=
{\upsilon\over |e_i|}$, $\rm vol = {\upsilon^3\over E}$, $E={ |e_1
e_2 e_3|\over \kappa}$, $e^D= |p|\sqrt{|m_0|/E} g$, $\upsilon =
\sqrt{|m_0|/E} r^2$, $\lambda=p^4|m_0|^{5/2} E^{-3/2}$. The
uplifting potential is positive and in these notation is given by
\be V_{up}\approx e^{K_{IIA}} V_0 \sim {g^4\over r^6}\ . \ee To see
if this makes the negative cosmological constant positive we would
prefer to deal with the potential depending on only one variable
like in the KKLT scenario where one can easily see that the AdS
minimum in $\sigma$ becomes a dS minimum meta-stable vacuum with the
barrier separating the minimum from the Minkowski vacuum at infinite
$\sigma$.  Here $\sigma$ is related to the total volume of the
compactified space. To proceed along these lines in type IIA model,
we would need to fix  the dilaton by a separate mechanism, make it
much heavier than the volume modulus, and integrate it out, as in
the IIB case. However, the quick glance on the mass matrix  in type
IIA theory presented in eq. (3.39) in \cite{DeWolfe:2005uu} shows
that no such situation can be easily found in the model
\cite{DeWolfe:2005uu}, since  the dilaton and the total volume are
fixed by the same combination of fluxes. In contrast, in type IIB
the dilaton is fixed by the electric and magnetic  3-form fluxes,
whereas the volume is fixed by the gaugino condensates and/or
Euclidean 3-brane instantons. Therefore the mass scales can be made
very different. It has been shown in \cite{Choi:2004sx} that in the
presence of the complex structure fields, such an hierarchy of mass
scales can be achieved, and the uplifting can be viewed as if it
takes place in the KKLT model with a single volume modulus.

Thus in the present situation we have to deal with the uplifting of
the minimum by using the  explicit expression for the potential for
the 2-moduli problem,
 \be {1\over \lambda} V_{total}\approx \Bigg[
{1\over 2} g^4 r^6 -\sqrt 2 g^3 + {1\over 4} {g^2\over r^6} +
{3\over 2} {g^4\over r^2}+ c{g^4\over r^6} \Bigg]. \ee Here the last
term in the potential is the uplifting term with $c={V_0\over
\lambda}>0$. First, we were using Mathematica to plot the potential
as a function of 2 variables. At $c=0$ it has a clear minimum with
the negative potential in the minimum. By gradually increasing the
uplifting coefficient $c$, we expected to achieve the uplifting to a
dS minimum. However, this was not happening: either there was a
minimum with negative cosmological constant, or no minimum
whatsoever.

In fact, we are able to prove a no-go theorem for the F-term upliting to de Sitter vacua in this model with arbitrary parameters. This explains the results of our negative Mathematica search but also proves in unambiguous way that there is no de Sitter vacua in this model.

We calculate the action of the differential operator $6 g
\partial_g - r\partial_r$  on the potential, following
\cite{DeWolfe:2005uu}, the difference being that we included the uplifting term. We find that at the minimum \be (6 g
\partial_g - r\partial_r)V_{total}= 18 V_{total}+ 12 \lambda \Bigg [
{g^4\over r^2}+ c {g^4\over r^6}\Bigg]=0 \ .\ee Therefore, despite
the attempts to uplift the potential, its value at the minimum of
the potential  remains negative, and there are no minima for the positive values
of the potential: \be V_{total}^{min}= -{2\over 3}\Bigg [ {g^2\over
r^6}+ c {g^4\over r^6}\Bigg]<0 \ . \ee
Thus even in presence of F-term uplifting  the model has only AdS minima.
\section{Discussion}

 One of the most interesting results of \cite{DeWolfe:2005uu} was a discovery of an infinitely large number of stable AdS vacua in  the type IIA massive string theory where the integer $N$ characterizing these vacua is unbounded. It has been argued in the literature and explained in the recent review \cite{Douglas:2006es}   that  the total number of  vacua  in this model is finite as a result of the physically  motivated cut-off of the large volumes of compactification.  But the main question remains whether any of these negative CC vacua could  be uplifted and become the meta-stable dS vacua with positive CC.

As we have shown, in the simplified version of the massive type IIA
model \cite{DeWolfe:2005uu} the F-term uplifting mechanism does not
work. One may try to improve this situation  by relaxing our first
assumption, i.e. by considering non-vanishing axions. However, since
the  K\"ahler potential does not depend on axions, it is unlikely
that the ``negative CC'' conclusion will change. One may also try to
relax our second assumption, and have the blow-up modes to change
the dynamics of the  stabilization significantly. Here it is easy to
get into a conflict with the supergravity approximation, but  it is
not a priory clear that this is impossible. This possibility requires
a separate investigation.
As we already pointed out, one may hope to find a satisfactory uplifting of the model in \cite{DeWolfe:2005uu} using the tools proposed in \cite{Saueressig:2005es}, \cite{Westphal:2005yz}.
Finally, one may consider a more general
class of models  based on   the type IIA  string theory and see
whether the F-term uplifting can be implemented in  any of these
models. However, our results demonstrate that it may  be a
challenging task  to uplift any of  the infinite number of the AdS
vacua found in \cite{DeWolfe:2005uu} and make them
phenomenologically acceptable. Additional studies of these issues
may affect our understanding of the structure of the landscape and
the criteria which should be applied towards the  counting of
quasi-realistic vacua.

\

\noindent{\large{\bf Acknowledgments}}

\noindent It is a pleasure to thank S. Kachru, A. Linde, M. Peskin,
S. Shenker and  L. Susskind  for stimulating discussions.   This
work  was supported by NSF grant PHY-0244728. MS is also supported
by the US Department of Energy under contract number
DE-AC02-76SF00515.

\providecommand{\href}[2]{#2}\end{document}